\documentclass[epj]{svjour}
\usepackage{graphics}
\begin{document}
\title{Coherent photon-photon interactions in
very peripheral relativistic heavy ion collisions}
\author{Gerhard Baur\inst{1,2} 
}                     
\offprints{Gerhard Baur}          
\institute{Institut f\"ur Kernphysik, Forschungszentrum J\"ulich,
D-52425 J\"ulich, Germany
\and
J\"ulich Centre for Hadron Physics, Forschungszentrum J\"ulich,
D-52425 J\"ulich, Germany}
\date{Received: date / Revised version: date}
%
\abstract{
Heavy ions at high velocities provide very strong
electromagnetic fields for a very short time. 
The main characteristics of ultraperipheral
relativistic heavy ion collisions are reviewed,
characteristic parameters are identified.
The main interest in ultraperipheral heavy ion 
collisions at relativistic ion colliders like the LHC
is the interactions of very high energy (equivalent) photons with 
the countermoving (equivalent) photons and hadrons (protons/ions).
The physics of these interactions is quite different from
and complementary to the physics of the strong fields achieved with
current and future lasers. 
} 
\maketitle
\section{Introduction}
\label{intro}
Strong fields are a probe for fundamental physics. 
Modern lasers provide very strong fields. In this context 
it is interesting to ask what is the characteristics 
of strong fields occurring in relativistic heavy ion collisions.
In this contribution the physics of ultraperipheral relativistic 
heavy ion collisions is described. 
The fields are extremely strong, typically much stronger 
than the Schwinger critical field. On the other hand 
these fields act only for an extremely short time.
There are comprehensive  recent reviews 
on ultraperipheral heavy ion collisions \cite{baltz08,bht07}. 
Nevertheless it is of interest to
give an overview of  key  physics aspects in a 
qualitative and 'context-setting' way.
The extremely short duration of these pulses leads 
to a clear difference to strong field physics in lasers.

Electric fields of heavy ions are rather strong,
a good measure is the quantity
$Z\alpha$, where $\alpha\equiv \frac{e^2}{\hbar c} \sim 1/137$ is the fine structure
constant. For Pb we have $Z=82$ and $Z\alpha \sim0.6$.
In slow ($v<<c$) heavy ion collisions
a united atom is formed for an appreciable time
where $Z_1 + Z_2 > Z_{\rm overcritical}$ is possible.
For $Z>Z_{\rm overcritical} \sim 173$ the $s_{1/2}$-state dives into the negative 
energy continuum. 
In this contribution the opposite limit 
$v \sim c$ and $\gamma =1/\sqrt{1-(v/c)^2}>>1$
is assumed.

In Sect. 2 the characteristics of the fast and strong
electromagnetic pulse are described as well as the transition
from fast to slow heavy ion collisions. 
This transition is controlled by a parameter
related to the Keldysh parameter in atomic physics. Then it
is shown how to describe theoretically one- and multi-photon 
processes. A very convenient tool to
describe one-photon exchange processes is the 
equivalent photon approximation. 
Multi-photon processes are treated in the Glauber approximation.
Typical examples of  $\gamma-$nucleus, 
$\gamma-$proton, and $\gamma- \gamma$ collisions 
at hadron colliders are described. In Sect. 4 
higher order processes are considered. Multiphoton processes
are especially prominent in electron-positron pair 
production. This is due to the low mass 
of the electron. We briefly discuss
bound-free pair production and $e^+ e^-$ multiple pair production.
A conclusion is given in Sect. 4. 
\section{Photoproduction in ultraperipheral hadron-hadron
collisions}
\label{sec:1}
\subsection{Characteristics of the electromagnetic pulse}
In Fig.1 an ultraperipheral collision with impact 
parameter b is shown schematically.
\begin{figure}
\resizebox{0.40\textwidth}{!}{
  \includegraphics{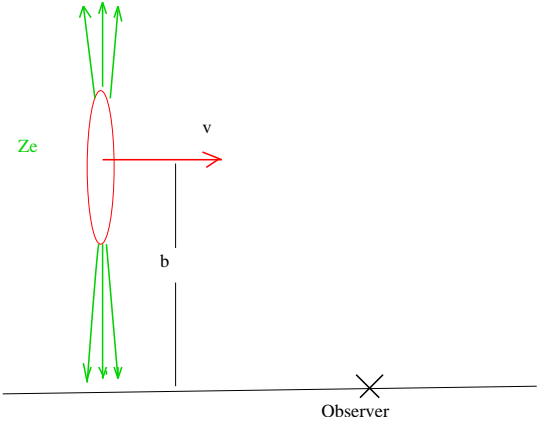}
}
\vspace{2cm}       
\caption{A schematic picture of an ultraperipheral collision.
The impact parameter b is larger
than the sum of the nuclear radii.}
\label{fig:1}       
\end{figure}
 The projectile
with electric charge $Ze$ has a velocity $v \sim c$.
A short and strong electromagnetic pulse is produced
with the following characteristics. Due to the 
Lorentz contraction the field is compressed, the maximum
field is in the direction perpendicular to the beam velocity
and is of the order of
$E_{\rm max} \sim \frac{Ze}{b^2} \gamma$, the magnetic field
is given by $\vec B_\perp=\vec v \times \vec E$.
The collision time is of the order of  
$\Delta t \sim \frac{b}{\gamma c}$ During this time, electrons 
can travel only  a 
distance of the order of  $\Delta l \sim c \Delta t=b/\gamma$. For impact parameters
corresponding to the electron Compton wavelength
$b \sim \hbar/(mc)$ this distance is  
$\Delta l \sim \hbar/(m c \gamma)<<386 \rm fm$. For many purposes, this pulse 
can be approximated by a $\delta$-function. 
In this case the perturbation series can be 
summed by exponentiation. One can neglect time ordering
and the S-matrix 
is given by $S=\exp{(i \int H_{int} dt/\hbar})$.

As a typical impact parameter we choose the Compton wavelength
of the electron. We obtain 
\begin{equation}
E_{\rm max}=E_{\rm critical} Z \alpha \gamma
\end{equation}
where $E_{\rm critical}=\frac{m^2 c^3}{\hbar e}
=1.3\times 10^{16} \frac{V}{cm}$ is the Schwinger
critical field strength. Since $Z \alpha \gamma>>1$ this field strength
is much higher than the Schwinger critical field.
The corresponding peak power is enormous,
it is of the order of 
$10^{29} (Z \alpha \gamma)^2 \frac{W}{cm^2}$,
see also \cite{meli}.
However, the correponding collision
time is extremely small, $\tau \sim \frac{\hbar}{m c^2 \gamma} 
\sim 10^{-21}\rm s/\gamma 
\equiv 1 \rm zs/\gamma$ 
(zs$=$zeptosecond). For RHIC (Au-Au, $\gamma=100)$
this is a small fraction of a zeptosecond,
for LHC (p-p, $\gamma=7000$ and Pb-Pb, $\gamma=3000$) 
this is a fraction of $10^{-24}\rm s \equiv 1 \rm ys$ (yoctosecond).
The classical momentum transfer on a test charge e is given
by $\Delta p \sim e E_{\rm max} \tau_{\rm collision} \sim
\frac{Ze^2}{bc}$, independent of $\gamma$. On the other hand,
the maximum energy of the equivalent photon spectrum 
rises linearly with $\gamma$: $E_{\gamma ,\rm max} \sim \frac{\hbar}
{\tau_{\rm collision}} \sim \frac{\hbar c}{b} \gamma$.

In \cite{landaf} the condition for the electric field to be 
considered as classical is given
by $ |\vec E|>>\sqrt{\hbar c}/(c \Delta t)^2$. Inserting the above 
estimates for the field and the collision time for an impact
parameter $b=\frac{\hbar}{mc}$ one obtains the 
condition $Z \sqrt{\alpha}>>\gamma$. For ultrarelativistic 
collisions ($\gamma > 10$) this cannot be fulfilled, i.e.
it is not useful to think of the field as a classical
quantity, it is best to think in terms of light quanta
(photons), as is usually done anyway.

\subsection{Transition from fast to slow collisions}
Electron-positron pair production in a constant electric 
field was studied by Schwinger \cite{schw}. Pair 
production becomes appreciable for a field strength
larger than the critical field strength $E_{\rm critical}= \frac{m^2c^3}{e \hbar}$.
The electron mass is denoted by $m$.
Electron pair production in the time-varying 
electric field $\vec E(t)= \vec E_0 \cos{\omega t}$ was studied in 
\cite{bi}. A  measure of adiabaticity is the parameter $\gamma_{\rm BI}
\equiv \frac{m c \omega}{e E_0}=\frac{\omega}{e E_0} \frac{\hbar}{\lambda_e}$ 
\cite{bi}. This parameter corresponds to 
the Keldysh parameter $\gamma_K=\frac{\omega}{e E_0}
\frac{\hbar}{a_{\rm Bohr}}$ in atomic physics. 
It was shown in \cite{bi} that
$\gamma_{\rm BI} <<1$ corresponds to the nonperturbative 
regime  (Schwinger  formula). For 
$\gamma_{\rm BI} >>1$ the process can be treated perturbatively.

The time- and space-dependence of the electromagnetic 
field in a heavy ion collision is more complicated
than the one envisaged in \cite{bi}.
Still, one can can find an
estimate of this parameter for fast heavy ion collisions.
As noted above, the maximum electric field strength is given by $E_{\rm max}= 
\frac{Z e }{b^2}\gamma$ where b is the impact parameter
and the field strength is appreciable for a time of the order
$\Delta t =1/\omega \sim \frac{b}{\gamma v}$. One finds
$\gamma_{\rm HI}=\frac{m c v b}{Z e^2}$. An appropriate
minimum impact parameter for processes involving
electrons is the Compton wavelength, i.e. $b=\frac{\hbar}{m c}$.
in this case one has   
$\gamma_{\rm HI}  =\frac{v}{ c Z \alpha}$. 
I conclude that relativistic ($v \sim c$) ultraperipheral reactions 
can be treated in perturbation theory. The condition is $\frac{v}{c}
>Z \alpha$. 

\subsection{One- and multi-photon processes, Glauber theory}

One-photon exchange processes are conveniently descibed
by the equivalent photon approximation (Weizs\"acker-
Williams method), see \cite{baltz08} and further
references given there. The cross section is obtained as a folding of the 
equivalent photon spectrum $n(\omega)$ and the 
elementary photo-cross section $\sigma_\gamma (\omega)$:
\begin{equation}
\sigma=\int \frac{d \omega}{\omega} n(\omega) \sigma_\gamma (\omega)
\end{equation}
The equivalent photon spectrum is given approximately by 
\begin{equation}
n(\omega) \sim Z^2 (\frac{c}{v})^2 \log{\frac{\gamma v}{\omega b_{\rm min}}}
\end{equation}
The charges of the Z protons in the nucleus act coherently,
this leads to a $Z^2$ factor. There are many soft photons
and the spectrum extends up to maximum 
photon energies up to $\omega_{\rm max}= \frac{\gamma v}{b_{\rm min}}$.
Beyond this energy there is an exponential decrease of the 
spectrum.
For processes involving electrons and positrons a minimum impact parameter
is given by $\lambda_e=\frac{\hbar}{m_e c}$, for other processes
an appropriate minimum impact parameter is provided by the nuclear 
radius R. For RHIC energies ($\gamma=100$), this maximum energy
is about 3 GeV, for LHC(Pb-Pb) energies ($\gamma = 3400$) it
is 100 GeV (in the c.m. (lab)-system).

In slow collisions ($v<<c$) the electric field acts for a long
time and higher order processes (many photon exchanges) are very 
well known, e.g. the excitation of rotational bands in low energy
heavy ion Coulomb exciation. The strength of the interaction 
decreases with increasing velocity like $1/v$. Nevertheless
many photons can be exchanged even in a relativistic 
heavy ion collision.

A theoretical tool to describe such many photon exchange 
processes is the Glauber approximation.
The scattering amplitude is given by
\begin{equation}
f_{fi, \rm Glauber}= \frac{i \pi}{k} \int d^2b \, exp(i \vec K \cdot
\vec b) <f|exp(i \chi(\vec b))|i>
\end{equation}
where $\vec K$ is the momentum transfer. The eikonal $\chi (b)$
takes care of all the elastic and inelastic processes. We have 
\begin{equation}
\chi (b)= \chi (b)_{\rm nuc}+ \chi (b)_C + \chi (b)_{e^+e^-} + \chi
(b)_{\rm GDR} + \chi(b)_V + ...
\end{equation}
The term $\chi_{\rm nuc}$ describes the effect coming from the 
nuclear interaction between the two ions. In a sharp cut-off
approximation it is given by $ \exp(i\chi(b)_{nuc})= \Theta (b-2R)$.
The term $\chi_C=2\eta log (kb)$ describes elastic Coulomb scattering.
The Coulomb parameter can be also be much larger than unity for 
relativistic collisions, i.e. at LHC(Pb-Pb) one has $\eta= 49$.
It means that many 'elastic photons' are exchanged.
The term $\chi (b)_{e^+e^-}$ describes electron- positron 
pair production. The excitation of the giant dipole resonance (GDR),
the production of vector mesons V ($V=\rho,\omega,\phi, 
J/\Psi,
\Upsilon,...$) is given by the corresponding Glauber phases.
In this way, many photon exchange processes, like GDR excitation
and vector meson production, etc. can be conveniently described
theoretically. 
\begin{figure}
\resizebox{0.35\textwidth}{!}{
  \includegraphics{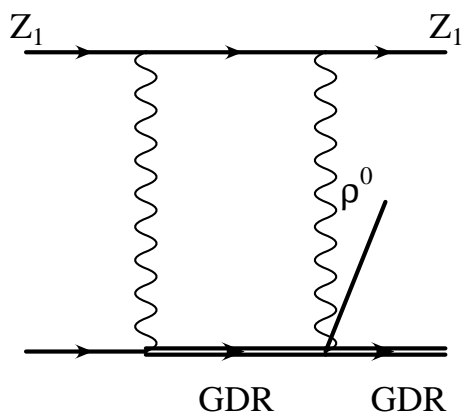}
}
\caption{A second order graph. The giant dipole 
resonance (GDR) is excited and a $\rho$-meson is produced
in a single collision }
\label{fig:2}       
\end{figure}
\begin{figure}
\resizebox{0.35\textwidth}{!}{
  \includegraphics{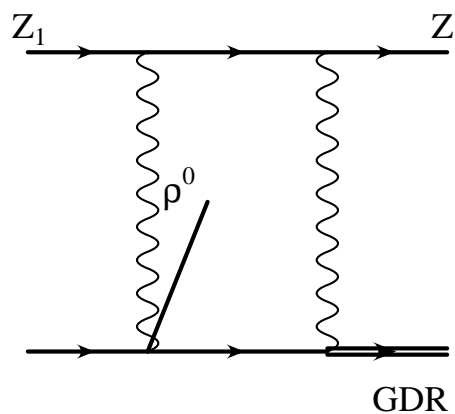}
}
\caption{Another second order graph.
The final state is the same as in Fig.2.}
\label{fig:3}       
\end{figure}
This is an efficient way to avoid the 
evaluation of Feynman diagrams in the limit of high
energies, see Figs. 2 and 3.
Further details can be found in \cite{npa03,hbt06}.
\subsection{$\gamma$-A, $\gamma$-p, and $\gamma- \gamma$ interactions
in ultraperipheral hadron-hadron collisions}
Even in the very fast relativistic 
heavy ion collisions multiphoton processes occur.
A good example is the excitation of 
the double phonon giant dipole resonance at GSI.
At GSI the Lorentz factor is of the order of $\gamma \sim 1-2$. 
The equivalent photon spectrum contains energies 
up to several tens of MeV one can, with
very large (of the order of one third) probability, excite the giant dipole
resonance at an excitation energy of 
$80A^{-1/3} \rm MeV$. Indeed, the excitation is so strong
that also a two-photon transition can occur to the 
higher harmonic mode, the double phonon giant dipole resonance
\cite{jim,schmidt,kon}.

This GDR excitation is also present at the highest 
energies, at RHIC (Au-Au) and LHC (Pb-Pb),
where it causes a beam loss. It also serves a useful purpose:
it is a trigger for UPC: in an ultraperipheral collision,
the giant dipole resonance is excited, which decays mainly by neutron
emission. These neutrons can be detected in the zero degree
calorimeter. 
At RHIC $\rho^0$-mesons were produced
in ultraperipheral collisions and detected with the STAR 
detector \cite{star}.
At LHC(Pb-Pb) the equivalent photon energy is
much higher than at RHIC and one expects to produce
and detect 
heavy vector mesons, possibly up to the $\Upsilon$ \cite{joakim}.
Vector meson photoproduction is a useful tool for QCD studies
\cite{baltz08,trento,cern}.

In AA collisions the photon spectrum is enhanced by a factor of $Z^2$,
on the other hand, the luminosity in pp collisions is appreciably
higher as compared to AA. This compensates to a large
extent, and also, due to the small size of the proton,
the equivalent photon spectrum extends to even higher energies.
At LHC(pp) electroweak and beyond the standard model 
physics studies are envisaged, see \cite{piotr}.

$\gamma - \gamma$ physics has been extensively studied at 
electron-positron colliders. $\gamma-\gamma$
collisions also occur in hadron-hadron
collisions. 
For heavy ions there is a $Z^4$-enhancement, on the other hand, the
beam luminosity is typically rather low.
The physics opportunities are described in \cite{baltz08},
unfortunately the cross section for the production of a 
Higgs boson in photon-photon fusion is too low to be observable.
In Fig. 4 a photon-photon collision is shown schematically.
Electron-positron pairs are copiously produced, of the 
order of 100 kb at LHC(Pb-Pb). The formation of positronium and muonium
is discussed in Ch.7.7 of \cite{bhtsk02}. 
Below the threshold for pair production there is 
only elastic scattering, with a very small cross section
with an $\omega ^6$ rise. Despite the high photon fluxes,
it seems hopeless to oberve this channel in heavy ion collisions.
\begin{figure}
\resizebox{0.35\textwidth}{!}{
  \includegraphics{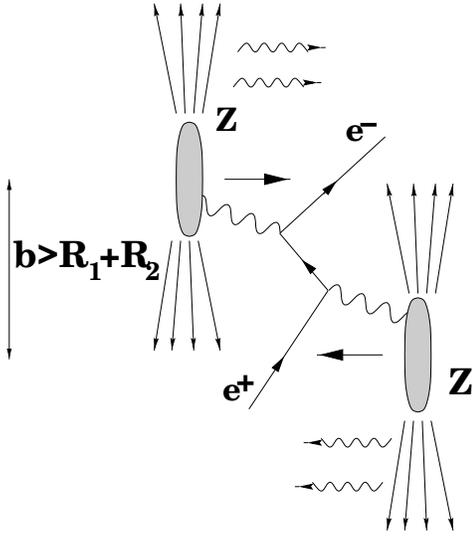}
}
\caption{A schematic picture of a photon-photon collision
in an ultraperipheral heavy ion collision.
Electron-positron pairs are produced with cross sections
of the order of 100kb}
\label{fig:4}       
\end{figure}
\section{Higher order effects}
In lowest order Born approximation electron-positron 
pair production is given by the Racah formula
\begin{equation}
\sigma_{\rm Racah}=\frac{28 \alpha^4 Z_1^2Z_2^2}{27\pi m^2}
(L^3-2.2L^2 + 3.84 L - 1.636).
\end{equation}
where $L\equiv \ln{\gamma^2}$. This pair 
production process is closely related to
the Bethe-Heitler process $\gamma + Z \rightarrow
e^+ + e^- + Z$, the real photon is now an equivalent photon, originating
from the other nucleus. 
Coulomb corrections to the Bethe-Heitler formula were
given by Bethe and Maximon. Quite analogously,
there are Coulomb corrections in the heavy ion case, see
\cite{iss} and the review \cite{bht07} where further 
references can be found.
Other higher order corrections are bound-free
pair production and multiple pair production,
which are briefly described in the following sections.
\subsection{Bound-free pair production}
Bound-free pair production is the process
\begin{equation}
Z + Z \rightarrow (Z+ e^-)_{\rm K-, L-,...shell} + e^+ + Z
\end{equation}
where the electron is produced
in a bound atomic orbit (K-,L-,..shell).
The cross section scales approximately as 
\begin{equation}
\sigma \sim \frac{Z^7 ln \gamma \delta_{l0}}{n^3}
\end{equation}
where n and l denote the principal and angular momentum
quantum numbers of the atomic bound state. 
These ions with their changed charge-to mass ratio will get
lost from the beam and they will heat up the beam pipe
in a hot spot.
It was
identified as a serious limit for the luminosity in Pb-Pb collisions at LHC,
due to the possible quenching of the superconducting magnets.
The bound-free pair production cross section 
was recently measured at RHIC \cite{bruce}, see
also 'The vacuum strikes back', Ref. \cite{pnu}.
Agreement with theory \cite{helmar} is good.
The bound-free pair production mechanism was also used in the production of 
fast antihydrogen in $\bar p -Xe$-collisions 
at LEAR in 1996 by W. Oelert et al.\cite{antihy}.
\subsection{Multiple pair production, a strong field effect}
The results of a lowest order calculation  \cite{htb05} of
electron-positron pair production probability $P^{(1)}$
are shown in Fig. 5. For small impact parameters
we have $P^{(1)}>1$ and unitarity is violated.
In order to restore unitarity, one has to consider the production of 
multiple pairs in a single collision
\cite{pra90}
\begin{equation}
Z + Z \rightarrow n (e^+ + e^-) + Z + Z .
\end{equation}
One may treat the  electron-positron pairs as quasibosons,
a coherent state \cite{glaube} of electron positron pairs 
is produced and   
one obtains a Poisson distribution for the production 
of N pairs P(N,b) : $P(N,b)=\frac{P^{(1)}(b)^N}{N!} \exp{-P^{(1)}(b)}$. 
This calculation bears some resemblance to the treatment
of the infrared catastrophe in QED.
\begin{figure}
\resizebox{0.45\textwidth}{!}{
  \includegraphics{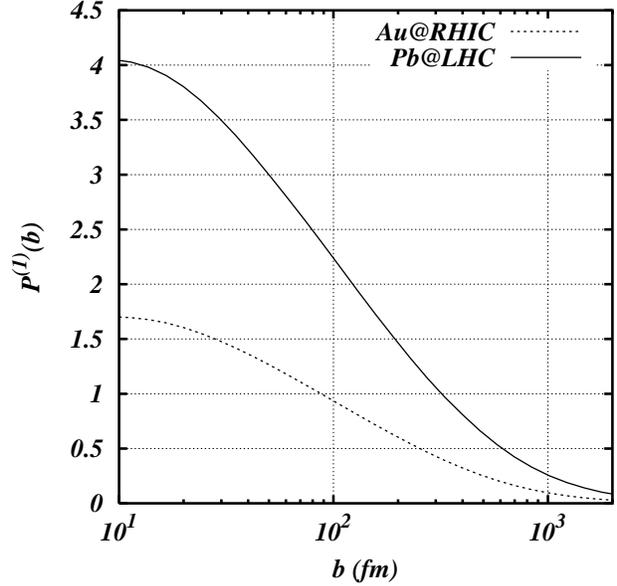}
}
\caption{Pair production probability
calculated in lowest order Born approximation.
Unitarity is violated manifestly. See text for restoration.}
\label{fig:5}       
\end{figure}
Multiple pair production is a strong field effect.
Although the cross section for multiple pair
production is quite high, of the order of kilobarns,
it will be  difficult to observe these processes at the LHC
due to the very low $p_\perp$ of most of the produced
electrons and positrons.
\section{Conclusion}
The electromagnetic fields occurring in relativistic
heavy ion collisions are extremely strong, they are
well beyond the Schwinger critical field strength.
However, these fields act only for an extremely short time.
Perturbation theory (Feynman graphs) is an appropriate tool to 
describe these processes.
Strong field effects manifest themselves in multi-photon
excitation processes, like multiple electron positron
pair production.
The main interest is in the  very high energy photons
which occur due to the extremely short interaction time.
This is in contrast to the physics of ultrahigh fields 
in laser physics.

%
%
\section{Acknowledgements}
I would like to thank Kai Hencken and Dirk Trautmann
for their collaboration over many years.
I am grateful to Dietrich Habs 
for interesting discussions and for 
inviting me to this stimulating workshop and school. 
I acknowledge the support by the European
Commission under contract ELI pp 212105 in the framework
of the program FP7 Infrastructures-2007-1.

%
%

\end{document}